\documentclass[aps,showpacs,showkeys,twocolumn]{revtex4}
\usepackage{mathrsfs}
\usepackage{amssymb}
\usepackage{amsmath}
\usepackage{bm}
\usepackage{graphics}
\usepackage{natbib}

\newcommand{\sss}[1]{{\scriptscriptstyle #1}}
\begin{document}
\title{Manipulability of the Kondo effect in a T-shaped triple-quantum-dot structure}

\author{Guang-Yu Yi$^1$}
\author{Cui Jiang$^3$}
\author{Lian-Lian Zhang$^1$}
\author{Su-Rui Zhong$^1$}
\author{Hao Chu$^2$}
\author{Wei-Jiang Gong$^1$}\email{gwj@mail.neu.edu.cn}

\affiliation{1. College of Sciences, Northeastern University, Shenyang 110819, China\\
2. Faculty of Robot Science and Engineering, Northeastern University, Shenyang 110819, China\\
3. Basic Department, Shenyang Institute of Engineering, Shenyang 110136, China}

\date{\today}

\begin{abstract}
We theoretically investigate the Kondo effect of a T-shaped triple-quantum-dot structure, by means of the numerical renormalization group method. It is found that the system's entropy and conductance have opportunities to exhibit abundant transition processes for different interdot-coupling manners, with the decrease of temperature. All these results are caused by the different Kondo physics mechanisms. Moreover, in the presence of appropriate parameters, the three-stage Kondo effect comes into being. Next when the electron-hole symmetry is broken or the structural parameters are changed, the Kondo resonance can also be observed in the conductance spectrum. However, it shows alternative dependence on the relevant quantities, i.e., the Coulomb interaction and interdot couplings. All these phenomena exhibit the interesting Kondo physics in this system. We believe that this work can be helpful for further understanding the Kondo effect in the triple-quantum-dot structures.
\end{abstract}
 \maketitle

\bigskip

\section{Introduction}
The Kondo effect, which first originates from the spin correlation between the impurity and conduction electrons of the host metal, results in the apparent increase of the resistance at sufficiently low temperature\cite{Hewson,Nagaoka,Wilson}. The underlying reason for this effect is explained as the spin-flip scattering processes of conduction electrons on the impurity spins\cite{Kondo,QFSun,Sasaki}. The successful fabrication of the quantum dot (QD) introduce new physics to the Kondo effect. When the Kondo QD is coupled to two leads, the conductance plateau appears in electron transport through the system, instead of resistance enhancement \cite{Goldhaber}. Compared with the conventional Kondo effect, multiple spin-flip processes take place for the conduction electrons, which give rise to the additional tunneling phenomenon. If the system is below the Kondo temperature, the resonant tunneling will be achieved in the case of the incident electron energy consistent with the Fermi energy of the whole system. Therefore, the conductance plateau arises in the transport spectrum with $\mathcal{G}_0={2e^2\over h}$. The Kondo effect in QD systems has already been one subject of extensive studies for more than two decades\cite{QD1,QD2,QD3,QD4,QD5,QD6,QD7,QD8,QD9,QD10}.

 \par
QD has its advantages that multiple ``atoms" can couple to one ``molecule" in different manners, and also, the metallic leads are allowed to connect with different QDs. This makes the Kondo effect more complex and exotic in QD-molecule systems \cite{DQD1,DQD2,DQD3,res1,res2,res3,res4,SU41,SU42,SU43,SU44,SU45,fano1,fano2,fano3,fano4,so1,so2,so3,so4,fer1}. During the past years, the Kondo effects in these systems have attracted extensive attentions. And lots of interesting phenomena have been observed. Take the double QD systems as an example, the Kondo effect can drive various results, including the transformation of the Kondo resonance\cite{res1,res2,res3,res4}, the $SU(4$) Fermi liquid behaviors\cite{SU41,SU42,SU43,SU44,SU45}, the Kondo-assisted interference effects\cite{fano1,fano2,fano3,fano4}, the interplay between the spin and orbital Kondo effects\cite{so1,so2,so3,so4}, the Kondo Effect influenced by the presence of ferromagnetic leads\cite{SU42,fer1}. Moreover, when the QD number in one molecule increases to three, the Kondo effect exhibit some special characteristics, such as the triple-anisotropic charge Kondo effect\cite{TQD1}, the ferromagnetic Kondo Effect\cite{TQD2}, and the symmetry-related Kondo effects\cite{TQD3}.
In view of the QD molecules, the T-shaped QDs are important geometries for the Kondo effect and its-related quantum transport behaviors. One typical phenomenon is the Fano-Kondo effect in the T-shaped double-QD structures, which has been observed on both the theoretical and experimental sides\cite{fano4,FanoK1}.
On the other hand, they are able to exhibit the two-stage Kondo effect, when the antiferromagnetic coupling between the QDs' spins competes with the Kondo effect\cite{fano4,tstage0,twostage1}. The two-stage nature of screening has been revealed especially in the temperature dependence of the conductance. With the decrease of temperature, for $T\sim T_K$, the conductance increases due to the occurrence of the first-stage Kondo effect, but when temperature further decreases to lower than the other characteristic temperature $T^{(1)}_K$, the conductance drops to zero, and the second-stage screening takes place.
\par
In the present work, we would like to investigate the Kondo effect of a T-shaped triple-QD (TTQD) structure. According to the previous works, the TTQDs exhibit some important geometry-related transport properties, especially when the strong correlation effects exist\cite{T3QD}. We then present the pictures of the Kondo physics in such a system from multiple aspects, by means of the numerical renormalization group (NRG) method. It is found that in this system, the Kondo resonance is allowed to occur in different energy regions, with the adjustment of the structural parameters. The underlying reason lies in that the Kondo temperatures in respective regions are related to the parameters, i.e., the Coulomb interaction and interdot couplings, in different ways. It is worth noting that the three-stage Kondo effect takes its influence to the quantum transport through this system, whose characteristics are apparently different from the two-stage Kondo physics.
\par
The rest of this paper is organized as follows. In Sec. II, we introduce the model Hamiltonian of the system and the method of calculation. The numerical results are presented and discussed in Sec.III. Finally, we give the summary in Sec. IV.

\begin{figure}
\begin{center}\scalebox{0.31}{\includegraphics{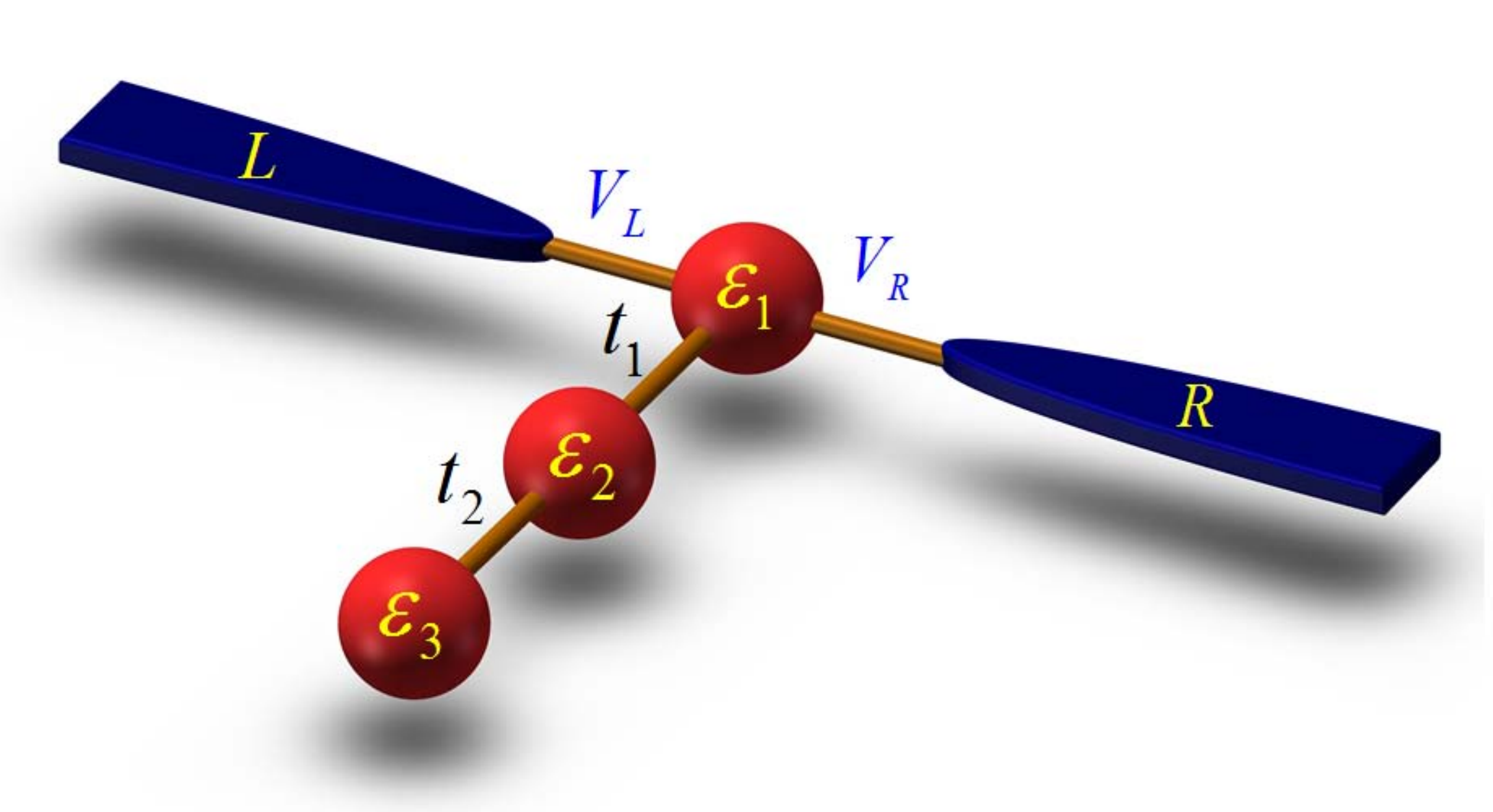}}
\caption{Schematic of the TTQD structure. QD-1 is coupled to two metallic leads with the coupling coefficients $V_L$ and $V_R$. The interdot couplings are $t_{1}$ and $t_{2}$, respectively.
\label{Struct}}
\end{center}
\end{figure}
\par

\par
\section{Model and theory}
The TTQD structure that we consider is illustrated in Fig.1, in which one terminal QD of the TTQDs (i.e., QD-1) connects with two normal metallic leads simultaneously. This system is described by the following Hamiltonian, i.e., $H=H_{C}+H_{QD}+H_T$. The first term is the Hamiltonian of the metallic leads, and it is written as
\begin{eqnarray}
H_C&=&\sum_{\alpha k \sigma} \varepsilon_{\alpha k}a^\dag_{\alpha k\sigma}a_{\alpha k\sigma}.
\end{eqnarray}
$a^\dag_{\alpha k\sigma}$($a_{\alpha k\sigma}$) is the operator that creates (annihilates) an electron with energy $\varepsilon_{\alpha k}$ for lead-$\alpha$ ($\alpha=L,R$), where $k$ is the momentum quantum number of the free conduction electron.
Next, $H_{QD}$ models the Hamiltonian for the TTQDs. We suppose that only one level exists in each QD with finite Coulomb interaction. It reads
\begin{eqnarray}
H_{QD}&=&\sum_{j\sigma}\varepsilon_{j} d^\dag_{j\sigma}d_{j\sigma}+\sum_{\sigma,j=1}^2t_{j} d^\dag_{j\sigma}d_{j+1\sigma}+h.c.\notag\\
&&+\sum_{j}U_{j}n_{j\uparrow}n_{j\downarrow}.
\end{eqnarray}
$d^\dag_{j\sigma}$($d_{j\sigma}$) is the operator to create (annihilate) an electron with energy $\varepsilon_j$ and spin $\sigma$ in QD-$j$ ($j=1,2$). $t_{j}$ is the interdot coupling coefficient, and $U_j$ indicates the strength of intradot Coulomb repulsion in the corresponding QD. The last term of $H$ denotes the coupling between QD-1 and the leads. For our considered system, it is directly written as
\begin{eqnarray}
H_T=\sum_{\alpha k\sigma}(V_{\alpha k}a^\dag_{\alpha k\sigma}d_{1\sigma}+h.c.).
\end{eqnarray}
$V_{\alpha k}$ describes the QD-lead coupling coefficient. In the wide-band limit with bandwidth being $W_D=2D$, the density of states can be viewed as $\rho_0={1\over2D}$. Accordingly, the coupling strength between QD-1 and leads can defined as $\Gamma_\alpha=\pi |V_\alpha|^2\rho_0$. This work only focuses on the case of symmetric QD-lead coupling with $\Gamma_\alpha=\Gamma$.
\par
Next in order to calculate the quantum transport properties governed by the electron correlation effects, we follow the theory pioneered by Meir and Wingreen\cite{Meir}. And then, the linear conductance in this system can be expressed as
\begin{equation}
{\cal G}(T)={\cal G}_0 \int^{+\infty}_{-\infty} \pi\Gamma (-{d f\over d\omega})\rho_d(\omega,T)d\omega,
\end{equation}
with ${\cal G}_0={2e^2\over h}$. $\rho_d=-{1\over\pi} {\rm Im}\sum_\sigma G_{dd,\sigma}$ is the local density of states (LDOS) in QD-1. $\mu$ is the system's chemical potential, and it can be supposed to be zero in the linear transport regime. In Eq.(4), we see that for studying the conductance, the key step is to solve the LDOS in QD-1. According to the previous works, the well-known NRG method is suitable to finish the relevant solution, when the system is in the Kondo regime\cite{NRG1,NRG2}. Thus, this work utilizes the full-density-matrix NRG method to study the conductance behaviors in our system governed by the Kondo physics. With respect to the NRG method, we should point out two points. On the one hand, in the iteration process, this work takes the logarithmic discretization parameter of the leads to be $\Lambda=4$, and keeps $6000$ states with the lowest energy in each iteration diagonalization step. For the lowest temperature in the iteration process, it is taken to be $T_{min}=10^{-24}D$ for calculation. On the other hand, to achieve the NRG method, such a system should be simplified to be a single-channel structure through the parity basis vector transformation\cite{NRG3}. In addition to the above two aspects, the $z$-average method is used to eliminate the parity error in the iteration process\cite{Zaverage}.

\par
Through the calculation with the NRG method, it is known that some fixed points of the group flow are related to different electronic states. The entropy $\mathcal{S}$ of the QDs of the whole system is determined by the microscopic states in the form of $\mathcal{S}\propto \ln W$ (The Boltzmann constant $k_B$ has been assumed to be 1). Following the diagonalization of NRG iteration, the system temperature decreases and the corresponding degree of freedom (microscopic states) decreases. Therefore, we can study the variation of the QD states by the change of the QD's entropy. Compared with the tunneling junction without QDs, the relationship between the contribution of QDs to the system's entropy and temperature can be expressed as
\begin{equation}
S_{\sss{d}}(T)={(E-F)\over T}-{(E-F)_0\over T}.
\end{equation}
Footmark 0 denotes the case without QDs, in which the energy of the system is also denoted as $E=\langle H\rangle={\rm Tr}(He^{-H/k_BT})$ with $F=-k_BT \ln{\rm Tr}(e^{-H/k_BT})$ being the system's free energy.
\begin{figure*}[htb]
\begin{center}\scalebox{0.17}{\includegraphics{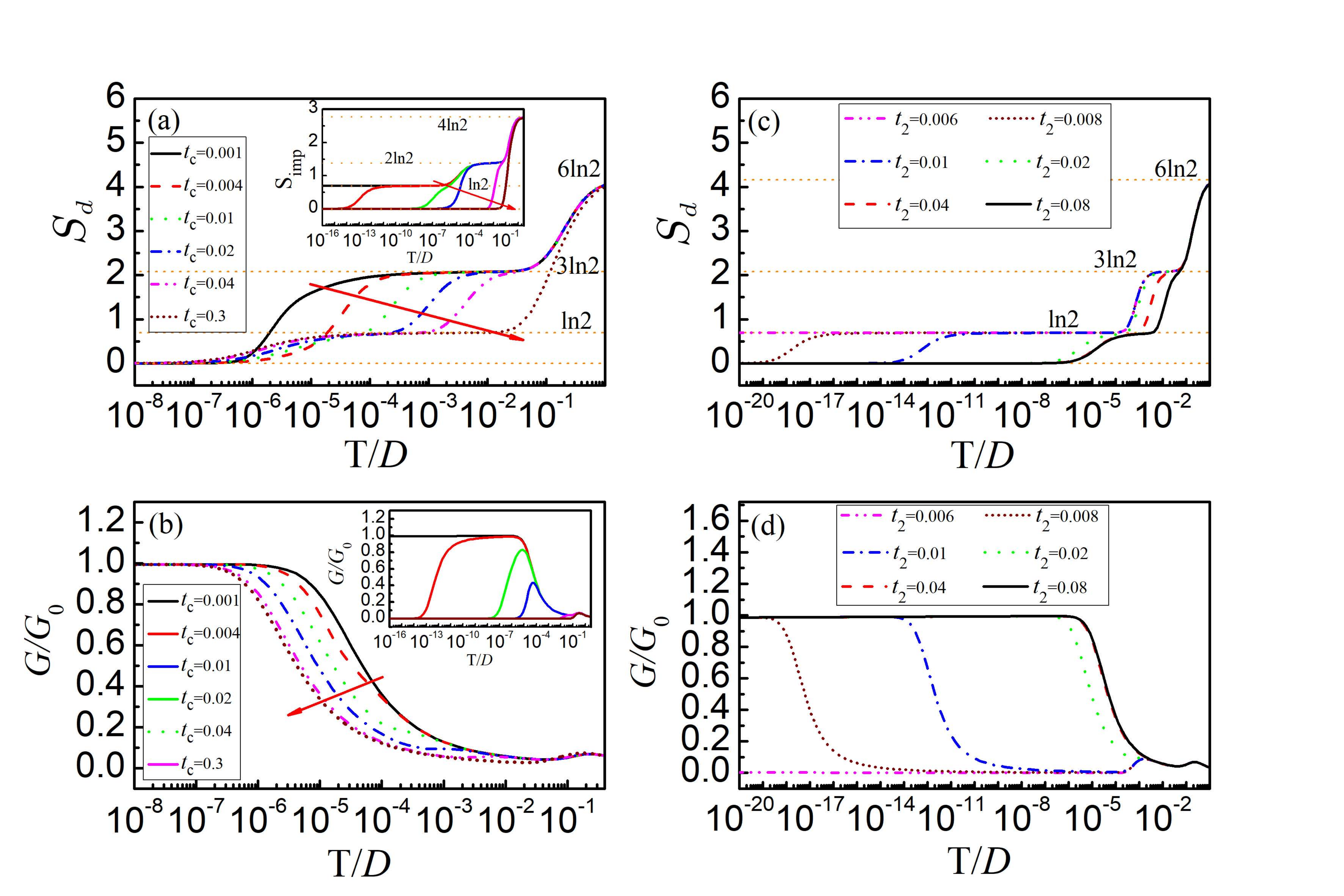}}
\caption{(a)-(b) Entropy and conductance as functions of temperature in the case of $t_j=t_c$. The insets are the corresponding results of the T-shaped double QDs. (c)-(d) Results of $t_1=0.02$, with the increase of $t_2$. The parameters are taken to be $U_j=0.8$ and $\varepsilon_j=-{U_j\over2}$.
\label{Phase1}}
\end{center}
\end{figure*}

\par
\section{Numerical results and discussions}
With the theory in the above section, we proceed to investigate the Kondo effects in our system, by calculating the linear conductance, entropy, and susceptibility. In the context, the half bandwidth $D$ is taken to be the energy unit, i.e., $D=1.0$, and the QD-lead coupling strength is $\Gamma=0.04$ (in unit of $D$). Such an assumption is reasonable according to the current experiments\cite{experiment}. Considering the structural complication of this system, we would like to perform discussion from two subsections.
\subsection{Exploring the three-stage Kondo effect}
\par
To begin with, we consider the case of $t_j=t_c$ to present the properties of entropy and conductance in the TTQD structure, as shown in Fig.2(a)-(b). We readily find the following three points.
\par
Firstly, when the interdot coupling is relatively small, e.g., $t_c=0.001$, the entropy transition shows platforms of $S_d=6\ln2$ and $3\ln2$ before arriving at 0, but no platform appears for $S_d=\ln2$. It is known that the platforms of $S_d=6\ln2$ and $3\ln2$ correspond to two different energy scales, respectively. The former is for the free-orbit regime (FOR), where the temperature is high, the QDs are independent and located at the FOR states. Namely, the four states $|0\rangle$, $|\uparrow\rangle$, $|\downarrow\rangle$, and $|\uparrow\downarrow\rangle$ in each QD are equiprobable, so the entropy approaches $S_{d}=6\ln2$. With the decrease of temperature, the system has an opportunity to enter the local-moment regime (LMR), in which the zero and double occupation are both forbidden. The two singly-occupied states in the system appear with equal probability, so the entropy is reduced by half to be $S_d=3\ln 2$. According to the previous works\cite{Zikton}, we can define the energy scale to describe the transition from $6\ln2$ to $3\ln2$, i.e., $T^*_1={U/\Delta}$, with $\Delta$ being a constant of the order one.
It can also be seen that in the limit of weak coupling, the system has a very wide temperature range with the entropy value at $3\ln2$. The reason is that the interdot coupling is too small, and then the spins in QD-2 and QD-3 is not easy to screen. Next when the temperature is low enough, e.g., $T\sim10^{-5}$, the entropy value begins to decrease to zero. This can be attributed to the independent formation of the Kondo singlet between QD-1 and the leads as well as the local spin singlet of QD-2 and QD-3. It should be noticed that since the local spin singlet forms in a long process, the entropy transition to zero is relatively slow.

\begin{figure*}[htb]
\begin{center}\scalebox{0.59}{\includegraphics{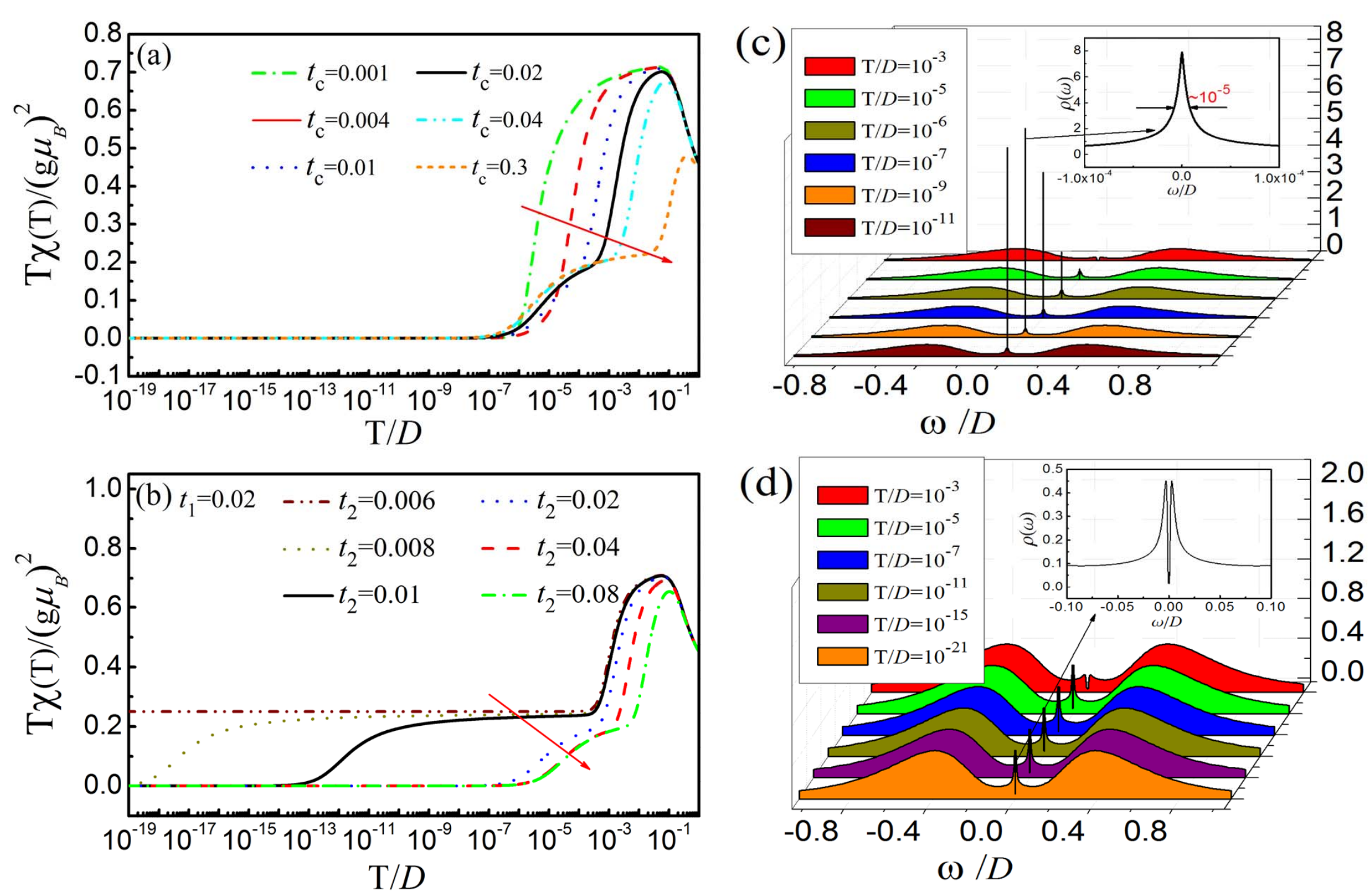}}
\caption{ (a)-(b) Temperature dependence of magnetic susceptibility in two interdot-coupling manners. (c)-(d) describe the LDOS spectra corresponding to the black lines in (a)-(b), respectively. Other parameters are identical with those in Fig.2. 
\label{a}}
\end{center}
\end{figure*}
\par
Secondly, for mediate interdot coupling, e.g., $t_c=0.02$, the system entropy experiences three platforms: $6\ln2$, $3\ln2$, and $\ln2$ before reaching zero. We see that the entropy transition from $6\ln2$ to $3\ln2$ are almost consistent with the above case, so they correspond to the same energy scale $T_1^*$. However, the following transition becomes dissimilar. This phenomenon should be attributed to the antiferromagnetic self-locking mechanism of the TTQD chain\cite{Ziktonn}. The three spins first lock into an antiferromagnetic spin chain at $T\sim T_{AF}$ and the collective spin with $\tilde{S}={1\over2}$ undergoes the Kondo screening at the lower temperature $T_K$. In such a case, we can estimate the value of $T_{AF}$ as $7\times 10^{-4}$ by analyzing the magnetic susceptibility when its value arrives at 0.25. It is clearly shown that at this position, the entropy value transits from $3\ln2$ to $\ln2$ [see Fig.2(a)]. After then, as temperature decreases, the entropy value decays to zero in the case of $T\sim 10^{-5}$. Such a temperature can directly be defined as the Kondo temperature $T_K$ for the mediate interdot coupling. Therefore in such a case, no characteristic temperature $T^{(1)}_K$ can be realized to drive the second-stage Kondo effect.

Thirdly, for the case of strong interdot coupling with $t_c=0.3$, the entropy value changes from $6\ln2$ to $\ln2$ directly, as shown in Fig.2(a). The underlying reason can be explained as the fact that $T_{AF}\sim T_1^*\gg T_K$ in this case. As a result, the Kondo temperature of the whole system is the transition temperature when the final residual entropy decreases to 0, just like the mediate interdot-coupling case. In fact, the TTQDs in such a case can be viewed as a large QD with effective spin $\tilde{S}={1\over2}$. Just for this reason, only one Kondo physics is involved with its characteristic temperature $T_K$.
\par

Next, we pay attention to the condition of $t_1\neq t_2$ to study the entropy flow and conductance, as shown in Fig.2(c)-(d). Although the entropy Fig.2(c) is similar to Fig.2(a), the corresponding conductances of them are different obviously, as denoted by the results in Fig.2(b) and Fig.2(d). It is found in Fig.2(d) that when $S_{d}=\ln2$, the conductance of the system is seriously suppressed, opposite to the result in Fig.2(b) which exhibits enhancement of the conductance with the decrease of temperature. This exactly indicates that in the two cases, the residual entropy with $S_d=\ln2$ contains different physical mechanisms. As is known, the result in Fig.2(a) is caused by the competition between two energy scales i.e., $T_{AF}$ and $T_K$, but Fig.2(d) originates from the second-stage Kondo effect and leads to the suppression of the conductance. Therefore, we conclude that in Fig.2(c), the transition from $3\ln2$ to $\ln2$ corresponds to the characteristic temperature $T^{(1)}_K$ for the second-stage Kondo effect, and correspondingly, the process of entropy changing from $\ln2$ to 0 defines the lowest characteristic temperature $T^{(2)}_K$ for the third-stage Kondo effect. They give rise to the anomalous phenomenon of the conductance variation, namely, the third-stage Kondo effect restores the inhibition of the second-stage Kondo effect on conductance, as shown in Fig.2(d). This is important for the study of multi-stage Kondo effect. In addition, it can be found that little signal of the first-stage Kondo effect comes into being. The reason is that in the situation of $t_1\gg t_2$ and $t_1\sim \Gamma$, the coupling between QD-1 and QD-2 causes their spins to be screened almost at together. Thus, the transition process from $3\ln2$ to $\ln2$ is suppressed and the result of $T^{(1)}_K\sim T_K$ comes into being. As a result, no conductance enhancement process labeled by $T_K$ can be observed in Fig.2(c)-(d).

\par
In order to further explain the discussion on Fig.2, we plot the spectra of the magnetic susceptibility and LDOS in QD-1 in different interdot-coupling manners, as shown in Fig.3. First of all, in Fig.3(a) it shows that when $t_c=0.3$, the small platform of magnetic susceptibility, i.e., $T\chi/(g\mu_B)^2\approx0.25$, forms when temperature decreases to $10^{-1}$. This means that $T_{AF}$ for $t_c=0.3$ is larger than that of $t_c=0.02$, and then the TTQDs tend to be a large QD with effective spin $\tilde{S}={1\over2}$. On the other hand, in the case of $t_c=0.001$, the local antiferromagnetic correlation of QD-2 and QD-3 competes with the Kondo correlation between QD-1 and the leads. And then, no platform of $T\chi/(g\mu_B)^2\approx0.25$ can be observed. However in both cases, all degrees of freedom are frozen at the temperature $T\sim10^{-6}$, so the entropy and magnetic susceptibility decay to zero.
\par
In Fig.3(b) where $t_1=0.02$ with $t_2$ from $0.006$ to $0.08$, we would like to divide the magnetic susceptibility curve into two groups, i.e., $0.006\ge t_2\ge0.01$ and $0.02\ge t_2\ge0.08$, respectively. Their changes become two branches at $T\sim 10^{-4}$. The first branch experiences an obvious 0.25 platform, caused by the second-stage Kondo effect. The following decrease of the magnetic susceptibility should be attributed to occurrence of the third-stage Kondo effect, so the corresponding Kondo temperature is much lower than the case of $t_c=0.3$ in Fig.3(a) where the Kondo screening is contributed by the large QD molecule. The other group corresponds to the three curves that $t_2$ changes from $0.02$ to $0.08$, and their performance is approximately the same. When the temperature is $10^{-5}\sim10^{-6}$, they all decrease to zero. The essence is similar to that in Fig.3(a). Local antiferromagnetic correlation form between QD-2 and QD-3, and there is no possibility of two or three-stage Kondo effect. Therefore, only one Kondo physics occurs with its high Kondo temperature $T_K\sim 10^{-5}$.
\begin{figure*}[htb]
\begin{center}\scalebox{0.93}{\includegraphics{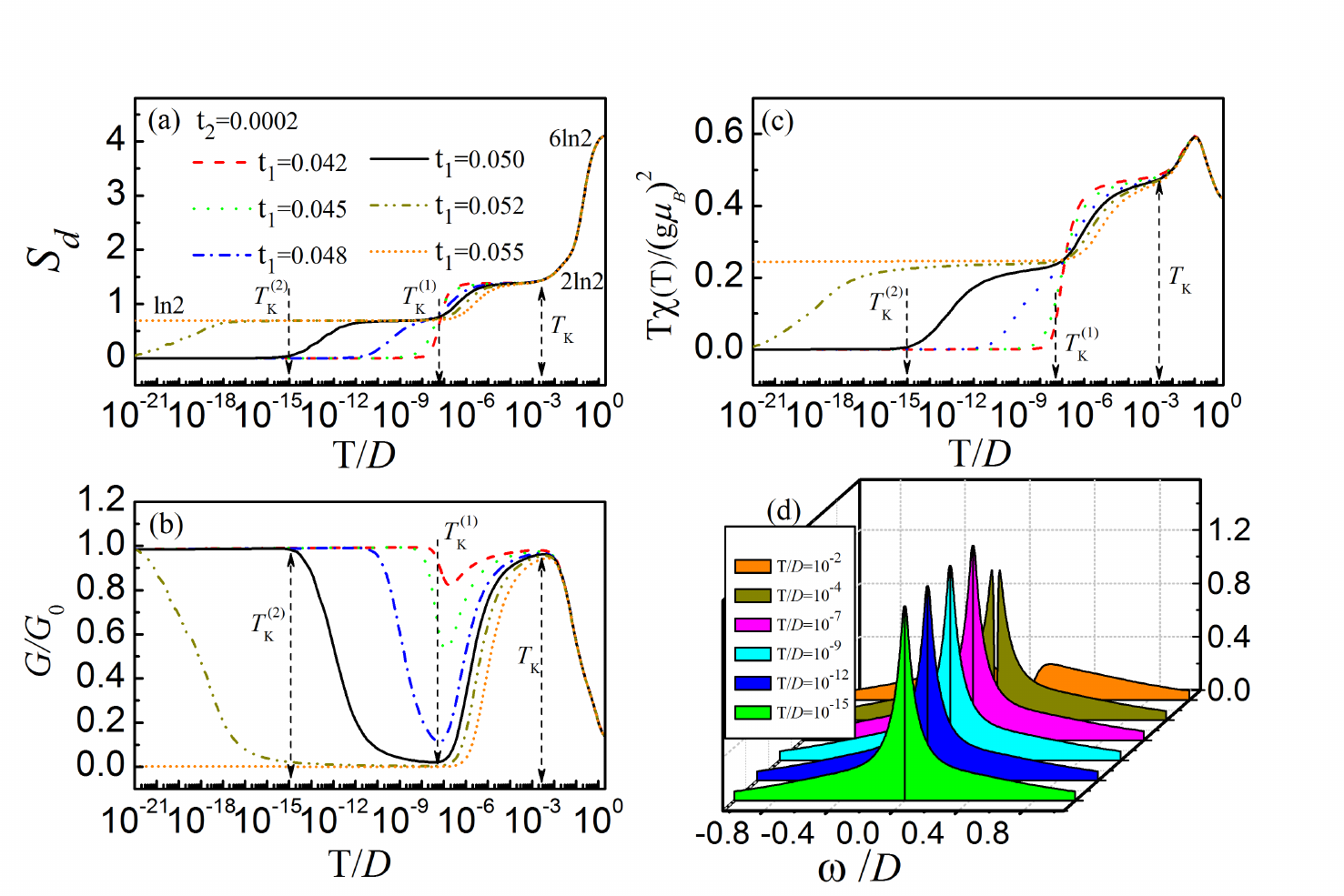}}
\caption{Spectra of entropy, conductance, magnetic susceptibility, and LDOS in QD-1, in the case of $t_2=0.0002$, with the increase of $t_1$. The parameters are taken to be $U_j=0.8$ and $\varepsilon_j={U_j\over2}$.}
\end{center}
\end{figure*}
\par
Following the magnetic susceptibility properties, we next investigate the LDOS in QD-1 to further uncover the underlying physics. The results are shown in Fig.3(c)-(d), which describe the cases of $t_c=0.02$ and $t_1=0.02$ with $t_2=0.01$, respectively. These two cases both show the platforms of magnetic susceptibility $T\chi/(g\mu_B)^2=0.25$ and entropy $S_d=\ln2$ [see Fig.2], but the LDOS spectra of QD-1 are quite different, as shown in Fig.3(c)-(d). For the former case, it is manifested as the normal Kondo effect. And with the decrease of temperature, the sharp Kondo peak appears at the energy zero point except from the Coulomb peaks of $\omega=\pm0.4$. In the inset of Fig.3(c), we measure the half width of the central peak and find that its value is consistent with the temperature estimated by the Schrieffer-Wolff transformation. Instead, in the latter case, due to the existence of the third-stage Kondo effect, the Kondo screening cloud spreads to QD-2 and QD-3 continuously, hence one can clearly find the splitting of LDOS peak around the energy zero point. The inset of Fig.3(d) shows that despite the decrease of temperature, the peak value does not change any more and keeps the same height.
Based on the above results, we know that the third-stage Kondo effect possesses strong electron correlation properties and low-temperature transport properties, which are completely different from the normal (first) and second-stage Kondo effects.
\begin{figure*}[htb]
\begin{center}\scalebox{0.09}{\includegraphics{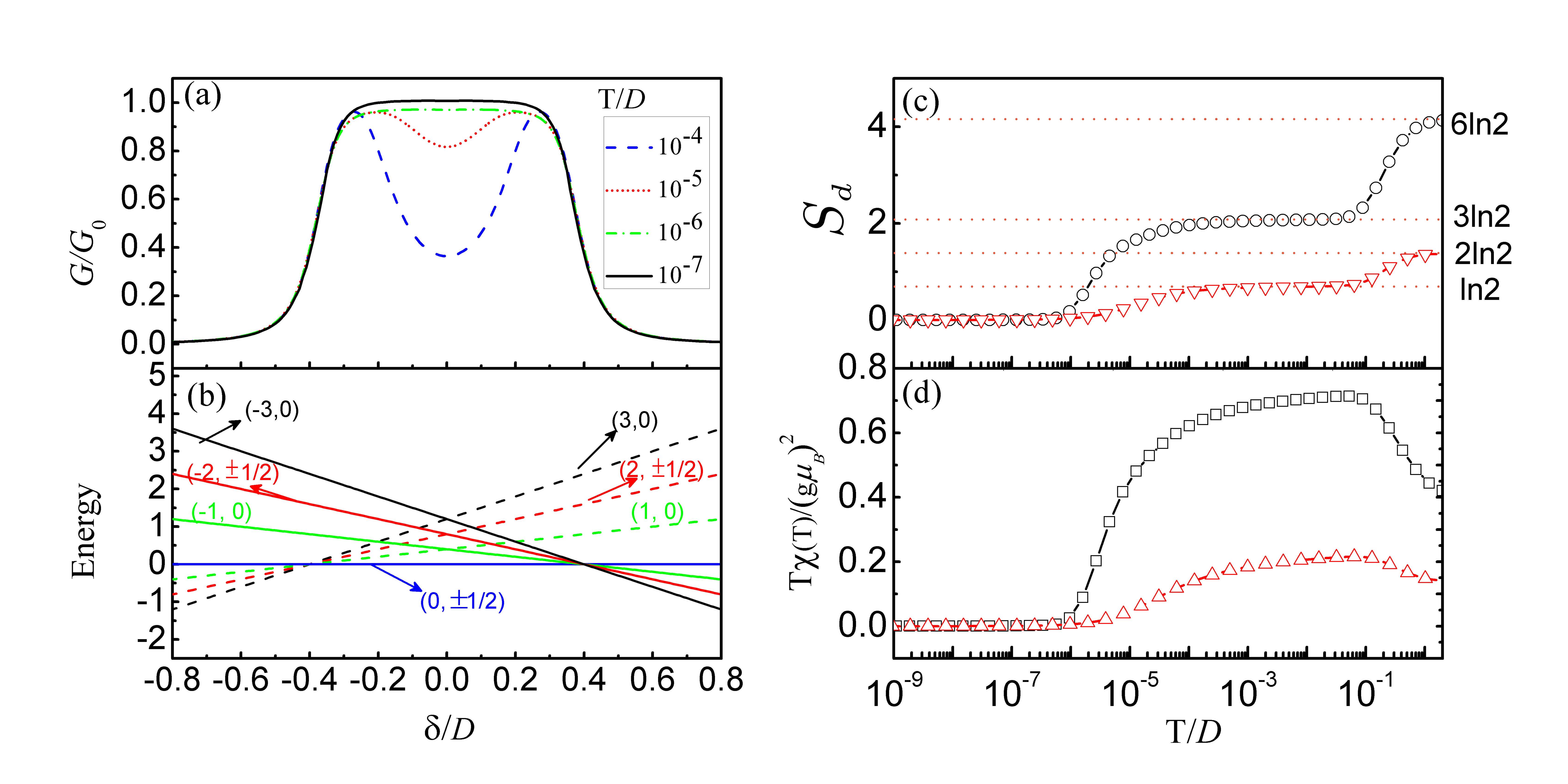}}
\caption{(a) Spectra of the linear conductance of the TTQDs with the decrease of temperature. (b) Ground-state energies in the respective subspaces as functions of $\delta$, for the isolated QD molecule. (c)-(d) Curves of entropy and magnetic susceptibility of TTQD system. Relevant parameters are taken to be $t_{j}=0.001$ and $U_{j}=0.8$.
\label{Phase1}}
\end{center}
\end{figure*}

\par
Next, we would like to think whether the first-, second-, and third-stage Kondo effect appear in the different cooling steps for the same set of parameters. The answer is clearly shown in Fig.4(a)-(d). Firstly, in Fig.4(a) it shows that if $t_2=0.0002$ with $t_1 \gg t_2$, a new entropy platform with $S_d=2\ln2$ appears. And when the entropy decreases from $6\ln2$ to $2\ln2$, the conductance increases to its unit value, as shown in Fig.4(b). We ascertain that the corresponding transition temperature is the first-stage Kondo temperature $T_K$. Next, as the temperature decreases, the entropy changes from the value of $2\ln2$ to $\ln2$, and the conductance value is suppressed. This should be described as the second-stage Kondo effect, with the corresponding transition temperature being $T^{(1)}_K$. As the temperature further decreases, the entropy decays to zero but the conductance magnitude is raised again, which is exactly caused by the third-stage Kondo effect. The transition temperature $T^{(2)}_K$ can be defined by observing the entropy value from $\ln2$ to 0. Consistent with the entropy and magnetic susceptibility, the magnetic susceptibility in Fig.4(c) also experiences three platforms of 0.5, 0.25 and 0, at the corresponding temperatures. Note, also, that the third-stage Kondo effect can be verified by Fig.4(d), because of the robust splitting of the LDOS peak of QD-1.
\par
The results in Fig.4 really show that if changing $t_1$, we can adjust the characteristic temperatures accordingly. This also means that by adjusting the structural parameters, respective Kondo physics pictures can be realized in the experiment. Take the green line in Fig.4 as an example, we obtain the result of $ T^{(2)}_K\sim10^{-8}$. If the value of $D$ is $10^5K$\cite{aaa}, there will be $T^{(2)}_K\sim1mK$ in practice. However, if we take the other set of parameters, e.g., $U=0.4$, $t_1=0.08$, $t_2=0.012$, and $\Gamma=0.17$, the calculated characteristic transition temperatures will increased to be $T^{(1)}_K\sim10^{-4}$ and $T^{(2)}_K\sim10^{-6}$, equivalent to $10K$ and $100 mK$, respectively. These results are practical for the current experiment conditions. According the relevant reports\cite{QD2,experiment}, lots of measurements about the Kondo effect in QD systems are performed in a refrigerator with a base temperature of $10 mK$. Therefore,  the respective Kondo temperatures in our system can be observed in experiment under the nowaday conditions for temperature.

\begin{figure*}[htb]
\begin{center}\scalebox{0.15}{\includegraphics{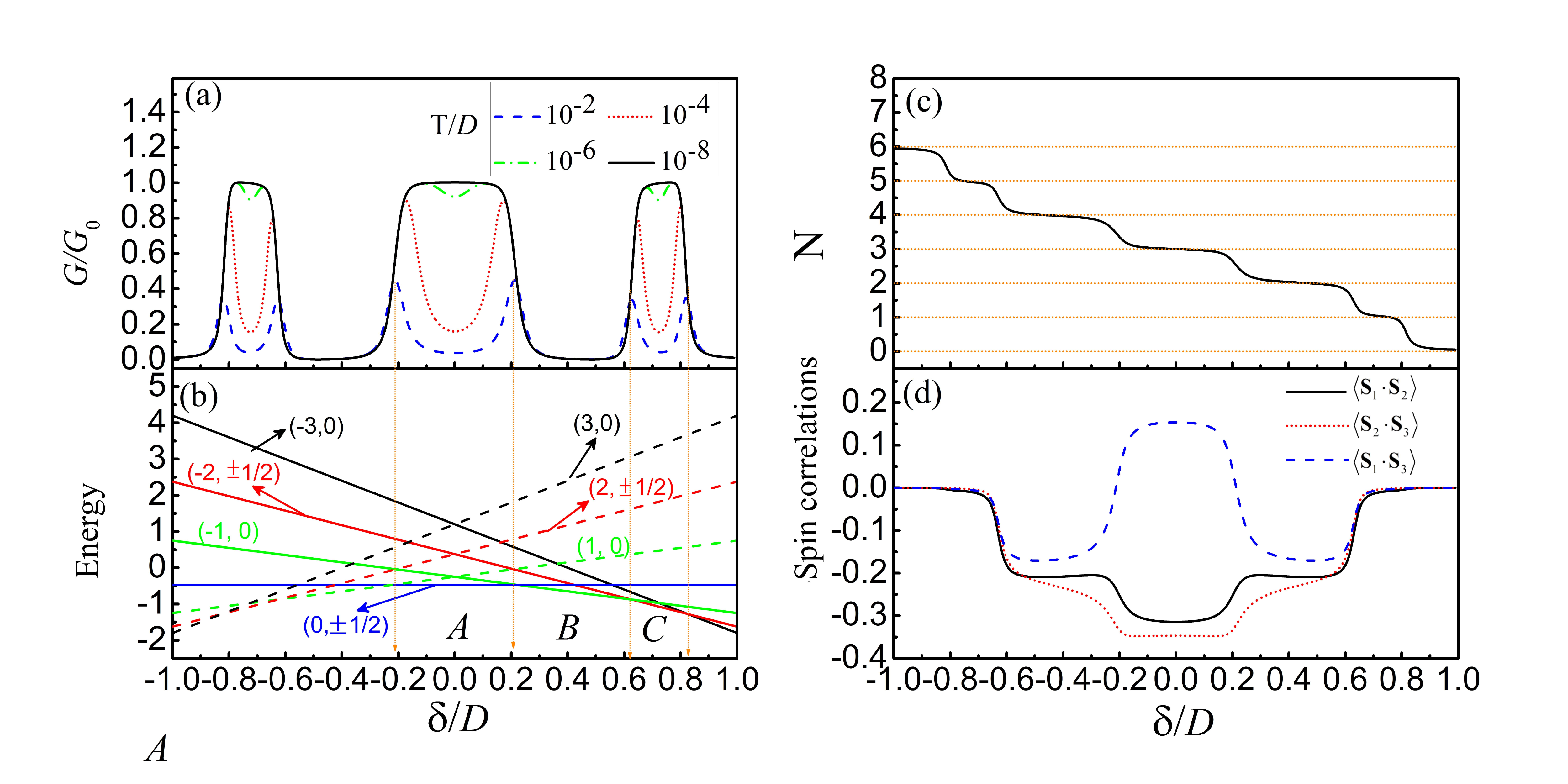}}
\caption{(a) Conductance spectra of the TTQDs with the decrease of temperature, in the case of $t_j=0.3$. (b) Corresponding ground-state energies in the different subspaces of the TTQDs. (c)-(d) Results of average electron occupation and interdot spin correlations. The intradot Coulomb interaction is fixed at $U_j=0.8$.
\label{Phase1}}
\end{center}
\end{figure*}
\par

\subsection{Further analysis about the Kondo physics}
\par
In the following, we performed detailed analysis about the Kondo effect by considering different cases. Fig.5 shows the results of weak interdot coupling where $t_j=0.001$. As for the QD levels and Coulomb interaction, we take $\varepsilon_j=\varepsilon$ and $U_j=U=0.8$. For benefiting our description, an additional quantity $\delta$ is introduced, defined as $\delta=\varepsilon+{U\over2}$. In Fig.5(a), we plot the conductance spectra of our system at different temperatures. Note that in the case of $\delta=0$, the Kondo temperature is $T^{}_K$. One can find in this figure that the conductance property is very similar to the experimental result in the single-QD case\cite{singledot}. When temperature is relatively high, e.g., $T\sim10^{-4}$, the system does not enter the Kondo regime. The conductance spectrum exhibits the Coulomb-blockade result between the two conductance peaks, where the electron transport is suppressed. As temperature gradually decreases, it has an opportunity to be below the Kondo temperature ($T^{}_K\sim 10^{-5}$), and then the conductance magnitude rises gradually, until the conductance plateau shows up as the unit value. It can be seen that the positions corresponding to the half-peak width of the conductance plateau is where $\delta=\pm0.4$, and it is exactly the position of $\pm {U\over2}$.
\par
Fig.5(b) shows the curves of the ground-state energies in the subspaces of the triple QDs without coupling to the leads, marked by the good quantum numbers $(Q, S_z)$. It is found that the positions of $\delta=\pm0.4$ are the crossing points of the ground states $(0,\pm{1\over2})$ and $(\pm3,0)$. It is clearly known that the Kondo physics and the corresponding transport behaviors in this case are the same as those in the single-QD result. However, the thermodynamic quantities, such as the entropy and magnetic susceptibility, are different, as shown in Fig.5(c)-(d). In Fig.5(c), we give the entropy curve of the single QD with the change of temperature (see the -$\nabla$- line). It shows that although the entropies of the TTQDs and the single QD are close to zero in the end, their high-temperature plateaus exhibit different values. This is exactly caused by the difference of the total spin between the TTQDs and single QD. Similarly, the magnetic susceptibility curves of the two systems in Fig.5(d) tend to zero from different values when the whole system arrives at the Kondo temperature. We then understand the difference between the weak-coupled TTQDs and the single QD cases.

\begin{figure*}[htb]
\begin{center}\scalebox{0.15}{\includegraphics{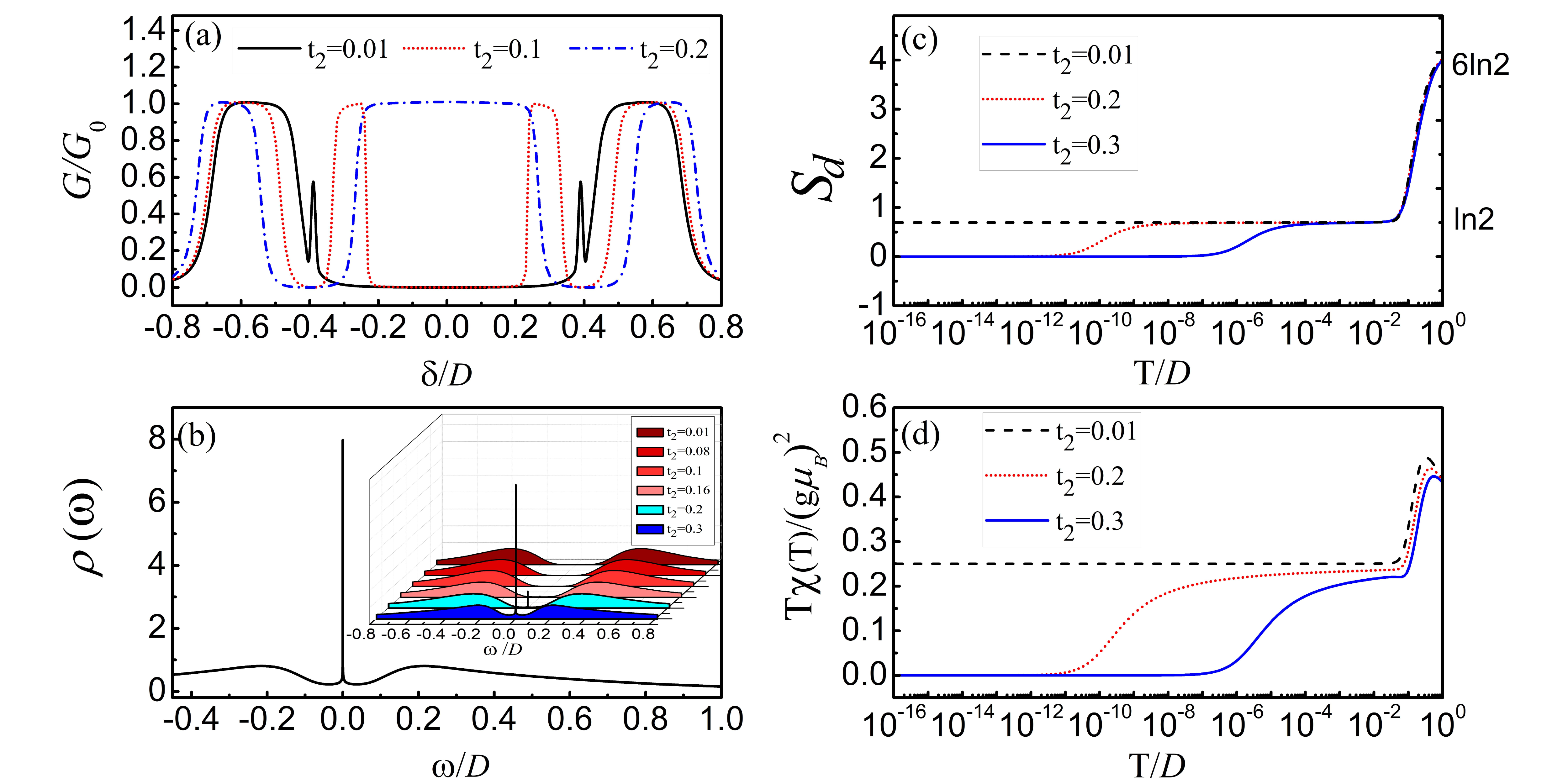}}
\caption{(a) Conductance in the cases of different $t_2$. (b) Density of states in QD-1 when $\delta=0$. (c)-(d) Entropy and magnetic susceptibility with the change of temperature.
\label{Phase1}}
\end{center}
\end{figure*}
\par
We next turn to the case of strong interdot coupling by taking $t_{j}=0.3$, and present the numerical results in Fig.6. Fig.6(a)-(d) are the spectra of the conductance, ground-state energies in respective subspaces, particle number, and spin correlations. Firstly, from Fig.6(a) we see that when the system is above the Kondo temperature, the Coulomb-blockade phenomenon is apparent, and three isolated Coulomb-blockade regions exist in the conductance spectrum. With the decrease of temperature, their Kondo transitions are almost the same. In the case of $T\sim 10^{-7}$, the Coulomb blockade changes to be the Kondo resonance, with the appearance of three Kondo plateaus. In the addition to the Kondo resonance around the electron-hole-symmetry point, the other Kondo resonances can be understood with the help of the results in Fig.6(b)-(c). One sees that as the temperature decreases to the Kondo temperature, the additional conductance plateaus appear in the positions where the particle number $N$ is equal to $1$ and $5$, respectively. The occupation of odd electron number is certainly causes the redundant spin to induce the Kondo effect. In Fig.6(d), it shows that when the system enters the Kondo regime, the TQDs construct one antiferromagnetic chain in the region of $N=3$. This verifies our description at the beginning of this section. Namely, in the strong-coupling case, the QD chain transforms into a large QD, and the whole system is equivalent to the Kondo model with its effective spin $\tilde{S}={1\over2}$.

\begin{figure}[htb]
\begin{center}\scalebox{0.55}{\includegraphics{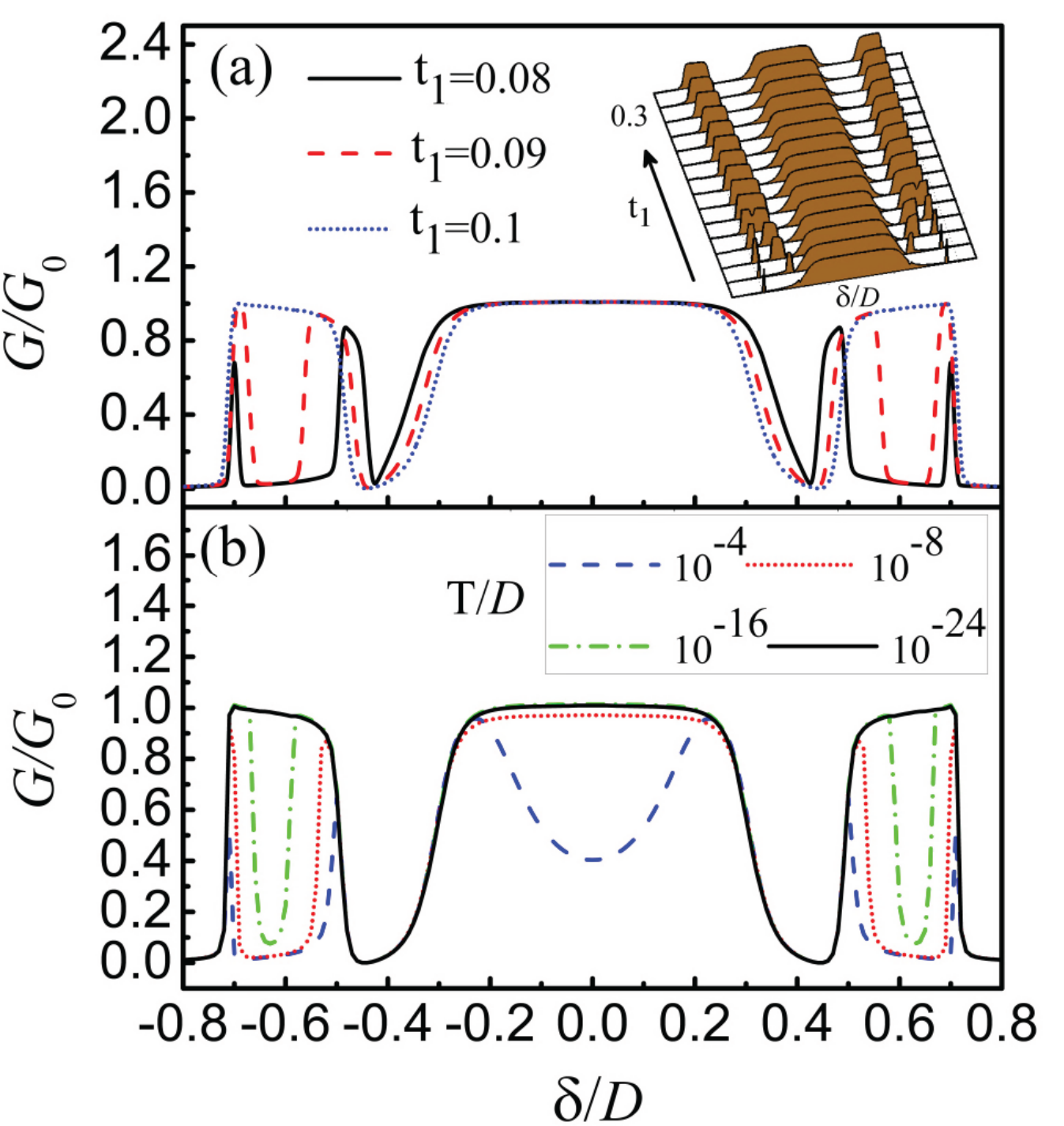}}
\caption{(a) Conductance spectra in the cases of different $t_1$. (b) Conductance vs $\delta$ for different temperatures with $t_1=0.1$. The interdot coupling $t_2$ is taken to be $t_{2}=0.3$.
\label{Phase1}}
\end{center}
\end{figure}
\par
After the discussions above, we would like to investigate the Kondo physics of our system in the case of $t_1\neq t_2$. Firstly, we take $t_1=0.3$ and change the value of $t_2$ to observe the transport behaviors. The numerical results are shown in Fig.7. It is found that for a relatively small $t_2$, e.g., $t_2=0.01$, a wide zero-value region appears in the vicinity of the electron-hole symmetry position, whereas two conductance plateaus exist near the points of $\delta=\pm0.5$. Thus, even when the second-stage Kondo effect takes place near the position of electron-hole symmetry, the Kondo resonance can survive in the other regions. Next, when $t_2=0.1$, two new conductance peaks appear near the position of $\delta=0$. And they are widened gradually with the further increase of $t_2$, until the formation of the conductance plateau, as shown in Fig.7(a). In Fig.7(b), we present the LDOS of QD-1 in the case of $t_1=0.3$ and $\delta=0$. The inset of Fig.7(b) describes the variation of the LDOS of QD-1 when $t_2$ increases gradually. It shows that for a small $t_2$, no Kondo peak emerge in the LDOS spectrum in the case of $\delta=0$. Under the strong-coupling condition, the two conductance peaks get close gradually, and then the local density of states exhibits the Kondo peak. The entropy and magnetic susceptibility curves in Fig.7(c)-(d) show that when $t_2$ is small enough, the Kondo temperature $T^{(2)}_K$ near the position of $\delta=0$ is very low. If it is lower than the lowest temperature $T_{min}$ in the NRG iteration process, the local magnetic moment will not be screened and the third-stage Kondo effect does not occur, thus the conductance will be very small. At the case of $t_2=0.3$, the Kondo temperature $T_K$ approaches $10^{-6}$, and the Kondo effect takes place, so the conductance plateau appears with its magnitude manifested as the unit value ${\cal G}_0$.

\par
In contrast with the case in Fig.7, we in Fig.8 take $t_2=0.3$ and change $t_1$ to present the conductance variation behaviors governed by the Kondo effect. It is found in Fig.8(a) that in this case, the conductance plateau is robust in the center of the conductance spectrum, and it is a bit narrowed with the increase of $t_1$. This can be attributed to the decrease of the Kondo temperature in this process, as discussed in Fig.2(b). However, for the cases of $\delta=\pm0.6$ [where $N=1(5)$], it shows that two conductance valleys are formed in the case of small $t_1$. Only when $t_1=0.1$, the conductance plateaus begin to come up. Besides, from the inset of Fig.8(a), we ascertain that as $t_1$ increases, the Coulomb-blockade phenomenon disappears gradually, in the low-energy and high-energy regions. As a result, three conductance plateaus exist in the conductance spectrum, and they are separate from each other by the conductance valley. Also, with the increase of $t_1$, the whole conductance spectrum is widened accordingly. In Fig.8(b), we present the conductance spectra in the case of $t_1=0.1$ and $t_2=0.3$. It clearly shows that in the case of small $t_1$, the Kondo resonance in the vicinity of $\delta=\pm0.6$ corresponds to a smaller $T_K$. For instance, when the system's temperature decreases to $10^{-24}$, the conductance plateau can also be observed due to the occurrence of the Kondo effect. So far, we have known the roles of $t_1$ and $t_2$ in adjusting the Kondo effect and its-related conductance properties.
\begin{figure}[h]
\begin{center}\scalebox{0.27}{\includegraphics{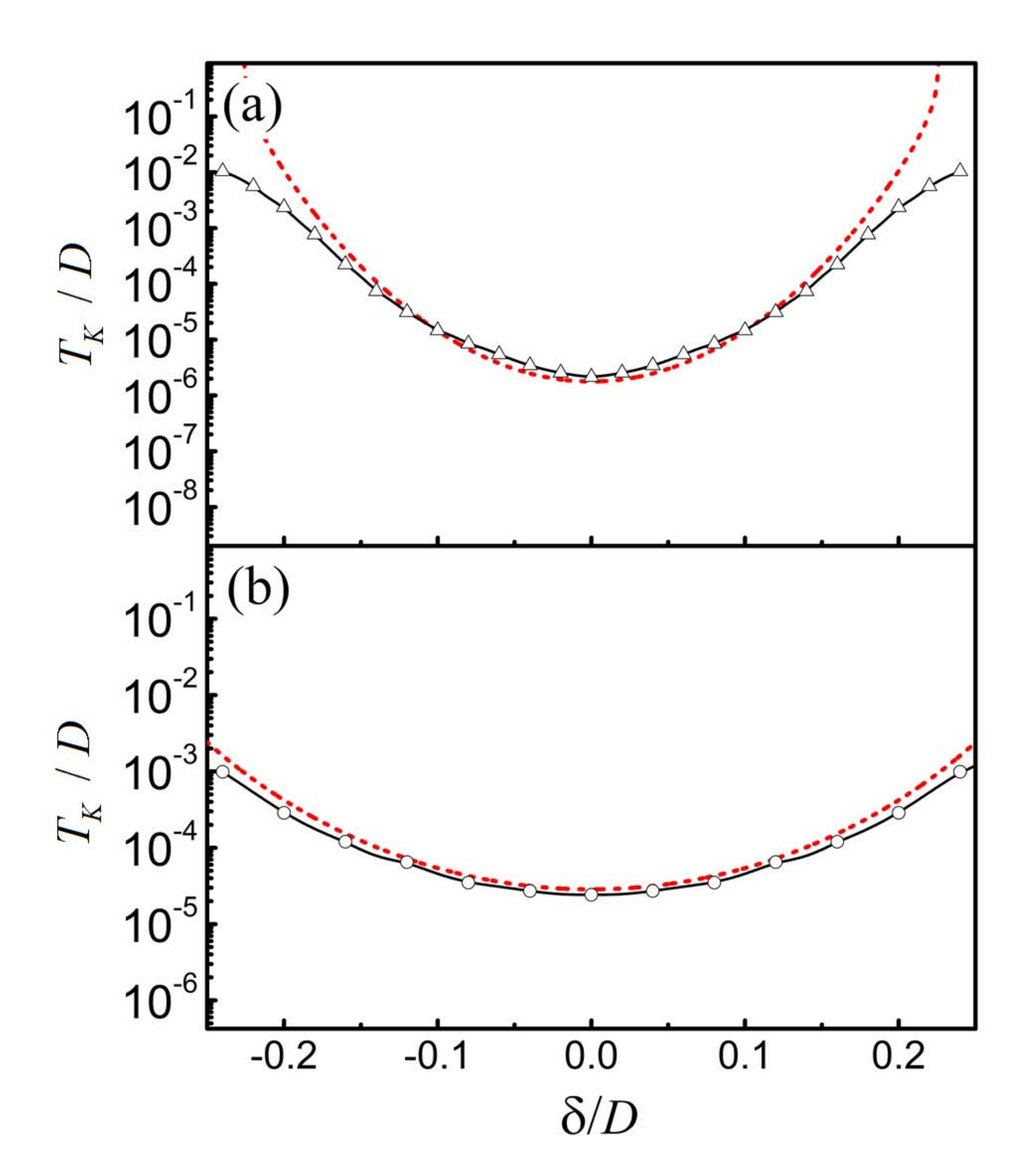}}
\caption{(a)-(b) Kondo temperature curves, corresponding to the cases of $N=3$ in Fig.6 and Fig.8, respectively. The dotted lines are obtained by Eq.(6). The Kondo temperature $T_K$ obtained by the \emph{entropy-temperature} relation are also labeled for comparison.
\label{Phase1}}
\end{center}
\end{figure}

\begin{figure}[h]
\begin{center}\scalebox{0.47}{\includegraphics{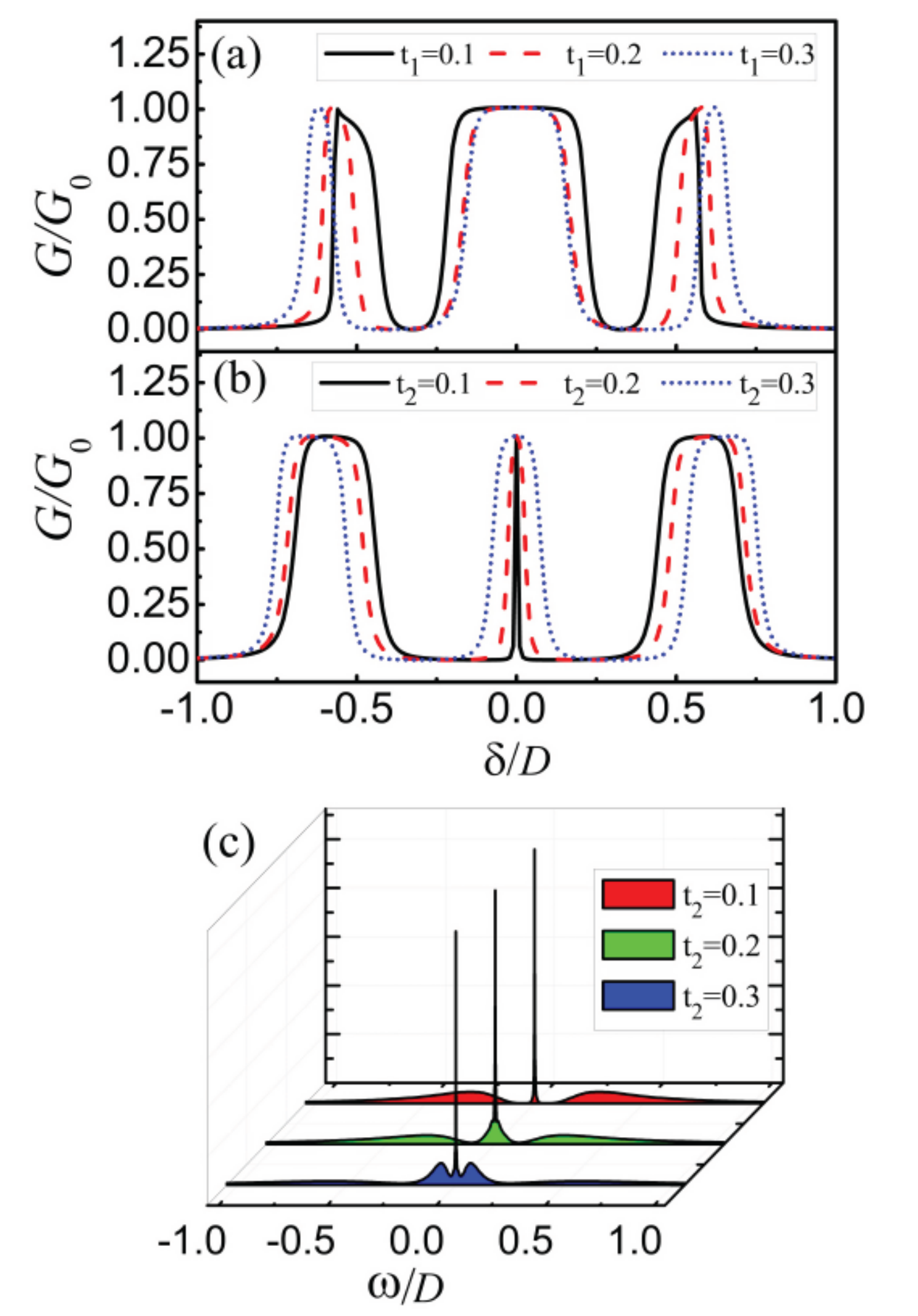}}
\caption{(a) Conductance influenced by $t_1$ in the case of $U_2=0$ and $t_2=0.3$. (b) Result of $U_3=0$ and $t_1=0.3$ for different $t_2$. (c) LDOS of QD-1 when $U_3=0$, $t_1=0.3$, and $\delta=0$.
\label{Phase1}}
\end{center}
\end{figure}

\begin{figure*}[htb]
\begin{center}\scalebox{0.33}{\includegraphics{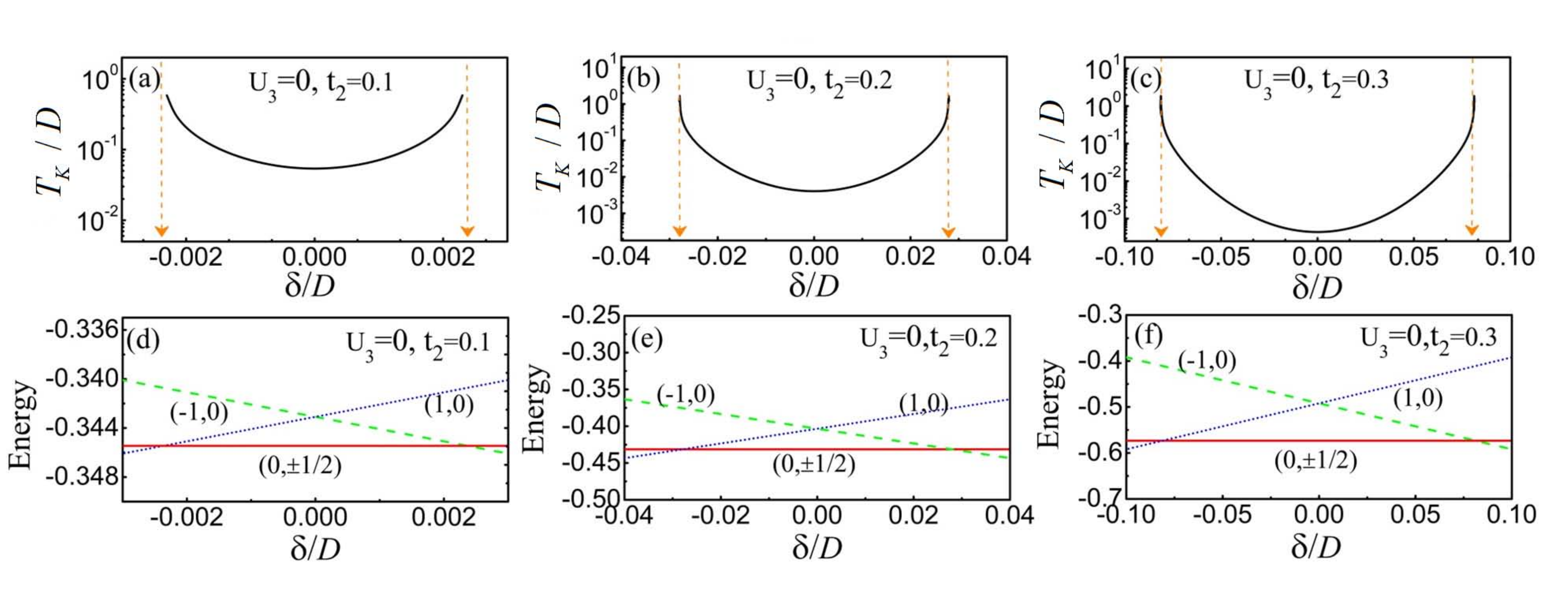}}
\caption{(a)-(c) Curves of the Kondo temperature $T_K$ near the point of $\delta=0$, in the case of $U_3=0$ and $t_1=0.3$. (d)-(f) The ground state energies in the special subspaces, corresponding to the cases of (a)-(c).
\label{Phase1}}
\end{center}
\end{figure*}
\par
In view of the above results, we next present the quantitative discussion about the Kondo temperature which is determined by the interdot couplings. To begin with, we employ the Schrieffer-Wolff transformation to solve the Kondo temperature, and the detailed process can be seen in the appendix. Through the complicated deductions, the Kondo temperature is estimated by the Haldane's expression\cite{SWT1,SWT2,SWT3}, i.e.,
\begin{equation}
T_K=0.182U\sqrt{J\rho_0}e^{-1/(J\rho_0)}
\end{equation}
where
\begin{small}
\begin{eqnarray}
J\rho_0&=&{2\Gamma\over\pi}(|\langle GS|d^\dag_{k\uparrow}|\mu\rangle|^2{1\over E_{GS}-E_\mu}\notag\\
&&+|\langle\nu|d^\dag_{k\downarrow}| GS\rangle|^2{1\over E_{GS}-E_\nu}).\notag
\end{eqnarray}
\end{small}
With the help of Eq.(6), we consider the case of $|\delta|\le 0.3$ where $N=3$ and present the curves of the Kondo temperatures corresponding to the central conductance plateaus in Fig.6 and Fig.8. The numerical results are shown in Fig.9(a)-(b), respectively. It can be seen from the figure that the Kondo temperature calculated by the theoretical formula (shown by dotted lines) is very close to the characteristic transition temperature (expressed by -$\Delta$- lines and -$\circ$-lines respectively) obtained from the measurement of entropy. The Kondo temperature is above $10^{-6}$, greater than the minimum temperature of the iteration diagonalization. So, the Kondo plateau comes into being. However, on both sides of $|\delta|>0.2$, the curves deviate from each other gradually. This is because the total number of particles in the system deviates from $N=3$ gradually, and the whole system can no longer be regarded as an effective Kondo model with effective spin $\tilde{S}={1\over2}$. As a consequence, the errors between them begin to take effect gradually.

\par
Finally, we try to consider the influence of different Coulomb interactions in the QDs on the Kondo effect, as shown in Fig.10. Fig.10(a) shows the conductance spectra of $t_2=0.3$ with different values of $t_1$. As for the Coulomb interactions in the QDs, we set $U_1=U_3=0.8$ and $U_2=0$. We find that no matter how $t_1$ changes, the conductance spectrum of the system shows three Kondo peaks. Compared with Fig.8(a), the difference between these two cases only lies in that increasing the values of $U_2$ lead to the narrowness of the insulating bands between two neighboring conductance plateaus. As for the interdot couplings, they play the opposite roles in modulating the conductance plateaus.
Inspired by Fig.10(a), we give the conductance spectra at different values of $t_2$ by setting $U_1=U_2=0.8$ and $U_3=0$ with $t_1=0.3$. The corresponding results are exhibited in Fig.10(b). It can be found that in such a case, the central insulating band shown in Fig.7 no longer appears, but a narrow conductance peak appears steadily, whose width is proportional to the value of $t_2$. Unlike Fig.7(b), the LDOS always always has a sharp peak despite the increase of $t_2$ [see Fig.10(c)]. And then, one stable conductance peak appears in Fig.10(b), for the case of small $t_2$.

In order to discuss the central conductance peak in Fig.10(b), in Fig.11(a)-(c) we present the dependence of the Kondo temperature $T_K$ on $\delta$ in the region of the half width of the peak. It can be seen that the half-peak width increases with the increment of $t_2$, but the Kondo temperature $T_K$ near the peak of $\delta=0$ decreases. This phenomenon can be understood by the results in Fig.11(d)-(f). When the number of particles is $N=3$, the lowest energy ground state is located in the subspace $(0,\pm{1\over2})$, and the Kondo temperature in Eq.(6) is only relevant to the imaginary excitation from states $(0,\pm{1\over2},0)$ to $(\pm1,0)$, i.e., $(\pm1,0,0)$. Therefore in Fig.11(d)-(f), the ground-state energy curves of these special subspaces are shown in the vicinity of $\delta=0$. It shows that the three curves in the figure can be enclosed in an isosceles triangle. As $t_2$ increases, the length of the bottom edge of the isosceles triangle increases gradually, and the vertexes correspond exactly to the positions shown by the two dotted arrows in Fig.11(a)-(c), which are exactly the positions of the half-peak width of the central conductance peak in Fig.10(b). After comparing the heights of the triangles formed by the ground state $(0,\pm{1\over2},0)$ and the state $(\pm{1},0,0)$, one can ascertain the energy of the virtual excitation for the Kondo effect, i.e. $\Delta E$. It shows that when $t_2$ increases from 0.1 to 0.3, $\Delta E$ increases from $2.365\times10^{-3}$ to $8.126\times10^{-2}$ monotonously. This change means the reduced probability of virtual excitation related to the Kondo effect. Accordingly in Fig.11(a)-(c), the Kondo temperature $T_K$ in the vicinity of $\delta=0$ decreases monotonously.

\section{Summary}
In summary, we have performed theoretical investigations about the Kondo effect of the TTQD structure, by means of the NRG method. As a result, it has been found that this system exhibits the intricate Kondo physics, because of the various structural parameters, e.g., the interdot coupling coefficients and intradot Coulomb interactions. One important result is that at the point of electron-hole symmetry, multiple-stage Kondo pictures have opportunities to come into being, which are related to the interdot coupling. Such a phenomenon is described by the different transition processes of the system's entropy, following the decrease of temperature. Via a comprehensive analysis, the pictures of the different-stage Kondo effect have been presented. In addition, it has been shown that when the system departs from the position of electron-hole symmetry, the Kondo resonance can also be observed, and it experiences alternative variation manners when the system's parameters are taken in different ways. Therefore in this TTQD structure, intricate Kondo physics is involved.
Based on these results, we think that this work can be helpful for further understanding the Kondo effect in the TTQD structures.

\section*{Acknowledgments}
\par
W. J. Gong thanks Tie-Feng Fang for his helpful discussions. This work was financially supported by the Fundamental Research
Funds for the Central Universities (Grant No. N2002005), the National Natural Science Foundation of China (Grants No. 11604221 and 11905027), and the LiaoNing Revitalization Talents Program (Grant No. XLYC1907033). Our numerical results are obtained via ``NRG Ljubljana"---open source numerical renormalization group code (http://nrgljubljana.ijs.si/).

\begin{appendix}
\section{Schrieffer-Wolff transformation}
For our system, it Hamiltonian can be rewritten as
$H=H_{0}+H_T$, with $H_0=H_C+H_{QD}$.
\par
In the Schrieffer-Wolff transformation, an effective Hamiltonian is given as
\begin{eqnarray}
H_{eff}&=&e^S He^{-S} \approx(H_0+H_T)+([S,H_0]+[S,H_T])\notag\\
&&+\frac{1}{2}[S,[S,H]]+\ldots.
\end{eqnarray}
To cut off the mixed terms, it is necessary to satisfy $H_T+[S,H_0]=0$.
With this condition, we obtain
\begin{eqnarray}
H_{eff}&=&H_0+\frac{1}{2}[S,H_T]+\frac{1}{2}[S,[S,H_T]]+\ldots\notag\\
&=&H_0+\frac{1}{\rm{2!}}[S,H_T]+\ldots.
\end{eqnarray}
For the part of $H_0$, there exists $H_0|m\rangle=E_m|m\rangle$, where $|m\rangle$ and $E_m$ are the eigenstate and eigenenergy, respectively. Thus,
\begin{align}
&\langle n|H_T|m\rangle=\langle n|SH_0|m\rangle-\langle n|HS_0|m\rangle\notag\\
&\rightarrow\langle n|H_T|m\rangle=E_{m}\langle n|S|m\rangle-E_{n}\langle n|S|m\rangle\notag\\
&\rightarrow\langle n|S|m\rangle=\frac{\langle n|H_T|m\rangle}{E_m-E_n}.
\end{align}
In the case of ${U\over\Gamma}\ll1$ and $\Gamma\ll|\varepsilon_{\pm}|$ ($\varepsilon_{\pm}=\varepsilon+{U\over2}\pm{U\over2}$),
\begin{small}
\begin{align}
&\langle n|H_{eff}|m\rangle\approx E_{m}\delta_{mn}+\frac{1}{2}\langle n|S,H_T]|m\rangle\notag\\
&=E_{m}\delta_{mn}+\frac{1}{2}(\langle n|SH_T|m\rangle-\langle n|H_TS|m\rangle\notag\\
&=E_{\alpha}\delta_{mn}+\frac{1}{2}\sum_\phi(\langle n|S|\phi\rangle\langle\phi|H_T|m\rangle-\langle n|H_T|\phi\rangle\langle\phi|S|m\rangle).\notag
\end{align}
\end{small}
Substituting Eq.(9) into the above formula, there will be
\begin{small}
\begin{eqnarray}
&&\langle n|H_{eff}|m\rangle\notag\\
&=&E_{m}\delta_{mn}+\frac{1}{2}\sum_\phi(\frac{\langle n|H_T|\phi\rangle\langle\phi|H_T|m\rangle}{E_m-E_\phi}\notag\\
&&-\frac{\langle n|H_T|\phi\rangle\langle\phi|H_T|m\rangle}{E_\phi-E_n})\notag\\
&=&E_{m}\delta_{mn}+\frac{1}{2}\sum_\phi\langle n|H_T|\phi\rangle\langle\phi|H_T|m\rangle\cdot\notag\\
&&(\frac{1}{E_m-E_\phi}+\frac{1}{E_n-E_\phi}).
\end{eqnarray}
\end{small}
\par
According to the existed research results, the charge fluctuation between QDs has no effect on the Kondo temperature. We can mainly consider the effective exchange interaction between QDs and the leads. For an isolated triple-QD system, there are $4^3$ eigenvectors, which can be divided into different subspaces according to the good quantum number $(Q, S_z)$. Suppose we study the $S={1\over2}$ Kondo effect near the position of $\delta=0$, which is half occupied with the ground-state spin being $\sigma={1\over2}$. The ground state can be expressed as $|0,\sigma,0\rangle$ (i.e., the ground state of the subspace with $Q=0$ and $\sigma={1\over2}$). We then concentrate on the virtual excitation from the global ground state to states $|\mp1,0,0\rangle$ which are the ground states in the subspaces $(\mp1,0)$. For the former case, the result in Eq.(10) is given as
\begin{align}
\frac{1}{2}\langle n|H_T|\mu\rangle\langle\mu|H_T|m\rangle(\frac{1}{E_m-E_\mu}+\frac{1}{E_n-E_\mu}),\notag
\end{align}
where $|\mu\rangle=|-1,0,0\rangle$.
$|m\rangle$ and $|n\rangle$ correspond to the ground state $|GS\rangle=|0,1/2\rangle$, whose energy is $E_{GS}$. And then, Eq.(10) has its more compact form, i.e.,
$|\langle GS|H_T|\mu\rangle|^2\frac{1}{E_{GS}-E_\mu}$.
\par
By a same token, in the latter case where $|\nu\rangle=|1,0,0\rangle$, we have $|GS|H_T|\nu\rangle|^2\frac{1}{E_{GS}-E_\nu}$.
Therefore,
\begin{eqnarray}
&&\langle n|H_{eff}|m\rangle=E_{m}\delta_{mn}+|\langle GS|H_T|\mu\rangle|^2\frac{1}{E_{GS}-E_\mu}\notag\\
&&+|\langle\nu|H_T|GS\rangle|^2\frac{1}{E_{GS}-E_\nu}.
\end{eqnarray}
Considering the expression of $H_T$, the Kondo expression of our system can be given as
\begin{align}
H_k=\sum_{k\sigma}Jc^\dagger_{k\sigma}\frac{\mathbf{\sigma}_{\sigma\sigma^{'}}}{2}c_{\kappa\sigma^{'}}\cdot \textbf{S},\notag
\end{align}
where
\begin{small}
\begin{eqnarray}
&&J=|V|^2\sum_{\sigma}(|\langle GS|d_{k\sigma}^\dagger|\mu\rangle|^2\frac{1}{E_{GS}-E_\mu}\notag\\
&&+|\langle\nu|d_{k\bar{\sigma}}^\dagger|GS\rangle|^2\frac{1}{E_{GS}-E_\nu})\notag.
\end{eqnarray}
\end{small}
After a simple deduction, one can find that $J\rho_0=\frac{2\Gamma}{\pi}(|\langle GS|d_{k\uparrow}^\dagger|\mu\rangle|^2\frac{1}{E_{GS}-E_\mu}+|\langle\nu|d_{k\downarrow}^\dagger|GS\rangle|^2\frac{1}{E_{GS}-E_\nu})$, with $\Gamma=\pi|V|^2\rho_0$.
Accordingly, the Kondo temperature in our system is expressed as
\begin{equation}
T_K=0.182U\sqrt{J\rho_0}\exp(-\frac{1}{J\rho_0}).
\end{equation}

\end{appendix}

\clearpage

\bigskip

\end{document}